\documentclass[journal=acsnano,manuscript=article,superscriptaddress,onecolumn,10pt]{achemso}
\usepackage[version=3]{mhchem} % Formula subscripts using \ce{}
\usepackage[section]{placeins}
\usepackage{rotating}
\usepackage{color}
\usepackage{array}
\usepackage{ulem}

\newcolumntype{R}{>{$}c<{$}}
\newcolumntype{L}{>{\centering\arraybackslash}m{0.12\linewidth}}
\SectionNumbersOn

\author{Luke Nambi Mohanam}
\affiliation{Department of Electrical and Computer Engineering, Boston University, Boston, MA 02215}
\author{Rafael Umeda}
\affiliation{Department of Physics, University of California, Irvine, Irvine, CA 92697}
\author{Lei Gu}
\affiliation{College of Physics and Electronic Engineering, Sichuan Normal University, Chengdu 610101, China}
\author{Yuanming Song}
\affiliation{Department of Chemistry, University of California, Irvine, Irvine, CA 92697}
\author{Doug J. Tobias}
\affiliation{Department of Chemistry, University of California, Irvine, Irvine, CA 92697}
\author{Allon Hochbaum}
\affiliation{Department of Materials Science and Engineering,University of California Irvine, Irvine, CA 92697, USA}
\alsoaffiliation{Department of Chemistry, University of California Irvine, Irvine, CA 92697, USA}
\alsoaffiliation{Department of Molecular Biology and Biochemistry, University of California Irvine, Irvine, CA 92697, USA}
\alsoaffiliation{Department of Chemical and Biomolecular Engineering, University of California Irvine, Irvine, CA 92697, USA}
\author{Ruqian Wu}
\affiliation{Department of Physics, University of California, Irvine, Irvine, CA 92697}
\email{wur@uci.edu}
\author{Sahar Sharifzadeh}
\affiliation{Department of Electrical and Computer Engineering, Boston University, Boston, MA 02215}
\alsoaffiliation{Division of Materials Science and Engineering, Boston University, Boston, MA 02215}
\email{ssharifz@bu.edu}

\title{Investigating electron conductivity regimes in the bacterial cytochrome wire OmcS}


\begin{document}

\begin{abstract}
The anaerobic bacterium \textit{Geobacter sulfurreducens} produces extracellular, electronically conductive cytochrome polymer wires that are conductive over micron length scales. Structure models from cryo-electron microscopy data show OmcS wires form a linear chain of hemes along the protein wire axis, which is proposed as the structural basis supporting their electronic properties. The geometric arrangement of heme along OmcS wires is conserved in many multiheme c-type cytochromes and other recently discovered microbial cytochrome wires. However, the mechanism by which this arrangement of heme molecules support electron transport through proteins and supramolecular heme wires is unclear. Here, we investigate the site energies, inter-heme coupling, and long-range electronic conductivity within OmcS. We introduce an approach to extract charge carrier site information directly from Kohn-Sham density functional theory, without employing projector schemes. We show that site and coupling energies are highly sensitive to changes in inter-heme geometry and the surrounding electrostatic environment\textcolor{black}{, as intuitively expected}. These parameters serve as inputs for a quantum charge carrier model that includes decoherence corrections with which we predict a diffusion coefficient comparable with other organic-based electronic materials. Based on these simulations, we propose that dynamic disorder\textcolor{black}{, particularly due to perturbative} inter-heme vibrations allow the carrier to overcome trapping due to the presence of static disorder \textit{via} small frequency-dependent fluctuations. These studies provide insights into molecular and electronic determinants of long-range electronic conductivity in microbial cytochrome wires and highlight design principles for bioinspired, heme-based conductive materials. 
\end{abstract}

\maketitle

%\section{Introduction}
In oxygen-free environments, microorganisms can survive and thrive \textit{via} respiratory metabolic pathways coupled to the reduction of extracellular metal species.~\cite{Lovley_1993} The reduction of extracellular soluble and insoluble metal species by microbes represents a key component of biogeochemical cycling~\cite{Nealson_1994} and bioremediation,~\cite{Kumar_2021,Watts_2012} bioenergy,\cite{WANG_2013,Logan_2012} and biosensing\cite{Atkinson_2022} technologies. The model bacterium \textit{Geobacter sulfurreducens} and other anaerobic microbes access extracellular electron acceptors, such as metals in the environment or poised oxidative electrodes in electrochemical cells, by passing spent electrons through outer membrane-associated \textit{c}-type cytochromes\cite{Lovley_1993,Edwards_2020,Gralnick_2023}. Cryo-electron microscopy (cryo-EM) analysis shows that \textit{G. sulfurreducens} assembles some of these cytochromes into conductive filamentous appendages, termed wires, aligning heme molecules in a one-dimensional chain along the wire axis and establishing a structural basis for long range electron transport over micrometer length scales.\cite{OmcS,OmcE,Wang_OmcZ,Gu_OmcZ} 

The arrangement of hemes in most multi-heme \textit{c}-type cytochromes can be classified as pairs stacked in a parallel manner, bringing the heme $\pi$-orbitals into close proximity, and as pairs with the heme planes set at obtuse angles -- so called “T-shaped” pairs.~\cite{vanWonderen_2019,OmcE} Except for two pairs sharing one highly solvent-exposed heme in OmcZ, the geometries of all heme pairs in \textit{G. sulfurreducens} OmcS, OmcE, and OmcZ wires, and in recently discovered archaeal extracellular cytochrome wires, cluster into closely spaced parallel and T-shaped arrangements.~\cite{Wang_OmcZ,Baquero_2023} A meta-analysis of heme pairs in the Protein Data Bank shows that the parallel and T-shaped arrangements are highly conserved across all deposited multi-heme \textit{c}-type cytochrome structures,\cite{Baquero_2023} suggesting a broader significance of these pairwise geometries in biological electron transfer. Nevertheless, the impact of these heme arrangements on long-range electron conduction in extracellular electron transfer cytochromes is poorly understood.  

 Electron conductivity within cytochromes is believed to be primarily diffusive,\cite{SmithJPCB_2006,Dahl_SciAdv2022, Blumberger_JPCL2020, second_heme_POD, first_heme_POD,Spin_Dependent_coupling,Guberman-Pfeffer_2022,Guberman-Pfeffer_2023} and inter-heme electronic coupling enables transport between charge carrier sites. Electronic coupling is determined by inter-heme distance and orientation, with greater $\pi$-orbital overlap leading to higher coupling,\cite{SmithJPCB_2006,Dahl_SciAdv2022,second_heme_POD, first_heme_POD,Guberman-Pfeffer_2022,Guberman-Pfeffer_2023}. Unlike electronic coupling, the measured relative diffusivity between different heme-proteins does not necessarily correlate with inter-heme orientation.\cite{Guberman-Pfeffer_2023} This is because varying local environments for hemes along the protein backbone causes site energy disorder, which results in carrier trapping.\cite{Blumberger_review_Marcus,second_heme_POD, first_heme_POD,Guberman-Pfeffer_2022,Guberman-Pfeffer_2023} In the frame of Marcus theory of electron transfer, this site energy disorder contributes to an energy barrier for electron transfer between heme molecules that must be overcome through thermal vibrations.\cite{Blumberger_review_Marcus} High level calculations based on Marcus theory produce a precise measure of this static disorder but still underestimate measured conductivity.\cite{Ing_2018,WANG_2013,Dahl_SciAdv2022,Wang_OmcZ,Guberman-Pfeffer_2022,Guberman-Pfeffer_2023} This discrepancy between theory and experiment can be reconciled if time-independent electronic coherence between hemes are incorporated empirically into the Marcus model.~\cite{Agam_delocalized_blocks,Eshel_delocalized_blocks} Recent applications of the Liouville-Master Equation, combined with Extended H{\"u}ckel method calculations for quantum states and the Landa{\"u}er-Buttiker theory, have been used to study the transport properties of OmcS and OmcZ.\cite{Papp_2024} Visualizations of transmission in real space demonstrated that extracellular cytochrome nanowires behave like insulating cables, with the inner core, spanning the chain of porphyrin rings within the proteins, contributing most to conductivity. This observation aligns with recent experimental findings that emphasize the importance of porphyrin rings within these filaments.\cite{Agam_delocalized_blocks} The development of complementary approaches for studying the transport properties of molecular wires is essential for the continued advancement of the field.
 
In this article, we utilize first-principles density functional theory (DFT) and a quantum conductivity model to investigate the mechanism of electron transport within the cytochrome wire OmcS. Our conductivity model is the mixed coherent-incoherent quantum conductivity model of Ru, Zhang, and Beratan~\cite{RZB_model}, with an empirical Lindblad correction to describe time-dependent decoherence, for an effective description of fast and non-adiabatic conduction with large static site energy disorder. We additionally introduce a procedure to account for non-perturbative electron-phonon interactions motivated by DFT studies. In order to parameterize this model from DFT, we introduce a method that is a computationally efficient approach to approximate site energies and inter-site couplings and is robust to the impact of nuclear motion. We show that while both coupling and site energies are sensitive to changes in the heme nuclear geometry, only site energies are sensitive to the electrostatic environment present in this computation. We find that with perfect inter-site coherence, the large static site energy differences within the OmcS chain cause low diffusivity due to Rabi oscillations. Non-instantaneous decoherence due to the Lindblad operator eliminates these oscillations, leading to orders of magnitude increase in diffusivity, while non-perturbative dynamic disorder introduced in the conductivity model smooths the energy landscape and allows the system to more easily overcome static site energy disorder, improving conductivity within certain parameter regimes. Overall, this study indicates that transient and incoherent atomic motion significantly impacts conductivity.

%This approach is an efficient and robust procedure to capture both quantum-mechanical (coherence) and bio-chemical (energetic fluctuations) features of the heme chains of cytochrome wires that facilitate long-range electron transport. 

\section{Results}
\subsection{First-principles DFT calculations of conductivity parameters}
In quantum transport theory, the conductivity of a molecular chain is determined by the dynamics of the wave function or equivalently the density matrix $\rho(t)$; specifically how rapidly and coherently a pulse-like quantum states can propagate across the chain. This process is influenced by quantum mechanical effects such as tunneling and decoherence. In this work, we adopt the simplified Lindblad master equation as discussed in Ref.~\citenum{RZB_model},

 \begin{equation}
     \frac{d\rho(t)}{dt}=\frac{-i}{\hbar} [\mathbf{H}, \rho ]+ L(\rho(t))
 \end{equation}, 
 
 Here the Hamiltonian, $\mathbf{H}$,  contains the site energies and nearest-neighbor couplings as,

 \begin{align}
   \mathbf{H} &= \sum_{N=0}^{N_{max}} 
    \varepsilon_N\left|N\right>\left<N\right| 
    %\nonumber\\ &+
   + \sum_{N=0}^{N_{max}-1} V_{N,N+1}
   \bigg(\left|N\right>\left<N+1\right| +
   \left|N+1\right>\left<N\right|
   \bigg),
    \label{eqn:H_lambda}
\end{align}

\noindent
where $N$ is the site index and $N_{max}$ is the number of sites in the model. $L$ is the Lindblad dephasing term, which physically represents the dynamic disorder for quantum states in the molecular chain, mostly via molecular thermal fluctuations, environmental coupling, and phonons. 

To parameterize Eq.~\ref{eqn:H_lambda} the charge carrier sites must be modeled by localized orbitals decoupled from their neighboring sites. Thus, Kohn-Sham (KS) DFT-computed orbitals, which correspond to adiabatic states, are generally projected onto orthogonalized diabatic states\cite{SmithJPCB_2006,FOD_paper,POD_paper}. While these projections have been successfully applied to many molecular systems including cytochromes,~\cite{Dahl_SciAdv2022,Blumberger_JPCL2020,second_heme_POD,first_heme_POD,PHE_POD_QM_MM,RZB_model,Guberman-Pfeffer_2022,Guberman-Pfeffer_2023} we take a different approach to obtaining heme site energies and inter-heme couplings in order to study trends with change in nuclear configuration, without being subject to \textcolor{black}{additional} errors associated with the orthogonalization procedure.~\cite{Bredas_2006,Blumberger_JPCL2020} 

\begin{figure}[htbp]
    \centering
    \includegraphics[width=0.49\textwidth]{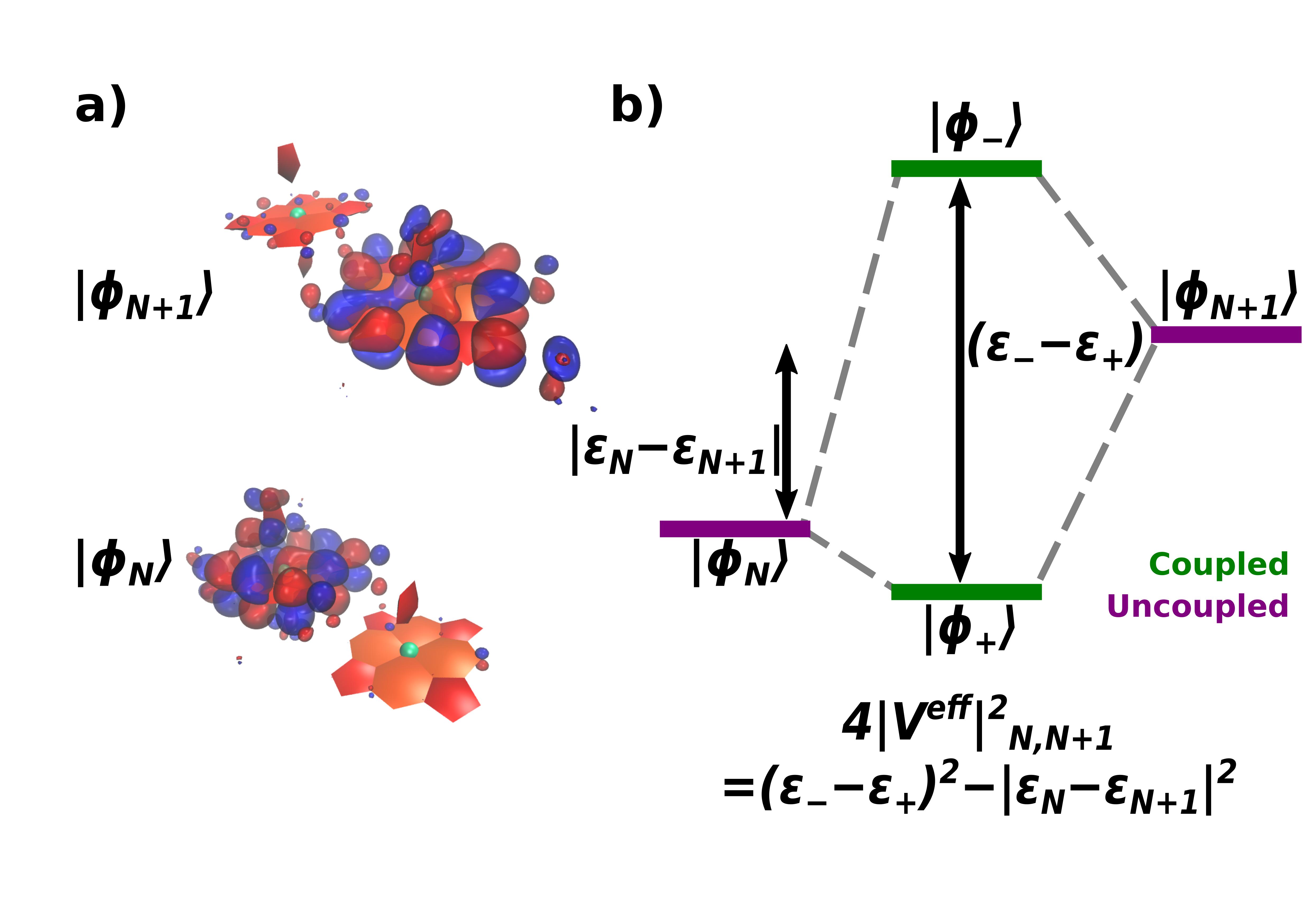}
    \caption{(a) Charge carrier sites on a heme pair. The electron density is centered around the Fe atom of the heme and delocalized over the porphyrin ring. (b) Schematic of of our approach for obtaining electronic coupling: The coupled orbitals are split in energy due to inter-heme electronic interactions and static disorder. By flipping the spin on one heme with respect to the other, the orbitals are decoupled. Site energies and effective electronic coupling are then calculated as shown in Eq.~\ref{eqn:veff}.}
    \label{fig:SFD}
\end{figure}

 Our approach (as illustrated in Figure~\ref{fig:SFD}) utilizes the fact that 1) the heme molecules associated with electron transfer contain unpaired spins, 2) unoccupied charge carrier sites with opposite spin are non-interacting, and 3) the total energy of the system is invariant with respect to the orientation of the unpaired spins, in order to extract the decoupled site energies, $\varepsilon_N$, and effective coupling, $ V_{N,N+1}$, directly from Kohn-Sham DFT. Consistent with previous studies~\cite{SmithJPCB_2006,Blumberger_review_Marcus,MtrX_review,Dahl_SciAdv2022,Blumberger_JPCL2020,second_heme_POD,first_heme_POD,PHE_POD_QM_MM,Spin_Dependent_coupling,Guberman-Pfeffer_2022,Guberman-Pfeffer_2023}, we assume that carriers conduct through one molecular orbital localized on each heme. Due to the orthogonality of spin components within spin-polarized unrestricted DFT, coupling between two heme molecules occurs only when orbitals on each site have parallel spins. By flipping the spin on one molecule with respect to the other, the Kohn-Sham DFT orbitals and energies are localized as shown in Figure~\ref{fig:SFD}a. 
  
Electronic overlap results in splitting of the orbital energies due to Pauli repulsion and correlation between electrons and so we can extract the coupling parameter V$^{eff}$ via the Kohn-Sham DFT-predicted orbital splitting.\cite{orbital_splitting_textbook_problem} However, because of the large static disorder among the heme molecules, there is an additional non-negligible orbital energy difference present. Here, the effective coupling for a pair of hemes is calculated as,\cite{orbital_splitting_textbook_problem}

\begin{equation}
|V^{eff}|_{N,N+1} =\frac{
\sqrt{\left((\varepsilon_- - \varepsilon_+)^{N,N+1}_{\uparrow,\uparrow}\right)^2 
- \left(|\varepsilon_N - \varepsilon_{N+1}|^{\,}_{\uparrow,\downarrow}\right)^2 }}{2}.
\label{eqn:veff}
\end{equation}

Here, $\uparrow$ and $\downarrow$ represent the \textcolor{black}{unpaired spin orientation} of each molecule, ($\varepsilon_- - \varepsilon_+)^{N,N+1}_{\uparrow,\uparrow}$ is the energy splitting associated with a coupled pair of molecules, and $|\varepsilon_N - \varepsilon_{N+1}|^{\,}_{\uparrow,\downarrow}$ is the site energy misalignment of the decoupled system. We utilize \textcolor{black}{decoupling via inter-heme spin flip} to obtain trends in coupling between heme pairs in different configurations. See SI Sections 1-3 for more details on \textcolor{black}{this approach}, its limitations, and  benchmarking of predicted coupling to a fragment orbital diabatization approach. \textcolor{black}{We also note that spin-based TDDFT approaches to determining coupling have been proposed previously.\cite{You2004_spin_flip,Gilbert2008_spin_flip,Hait2021_spin_flip,Corzo2022_spin_flip}} 

\begin{figure*}%[\sidecaptionrelwidth][t]
    \centering
    \includegraphics[width=0.7\textwidth]{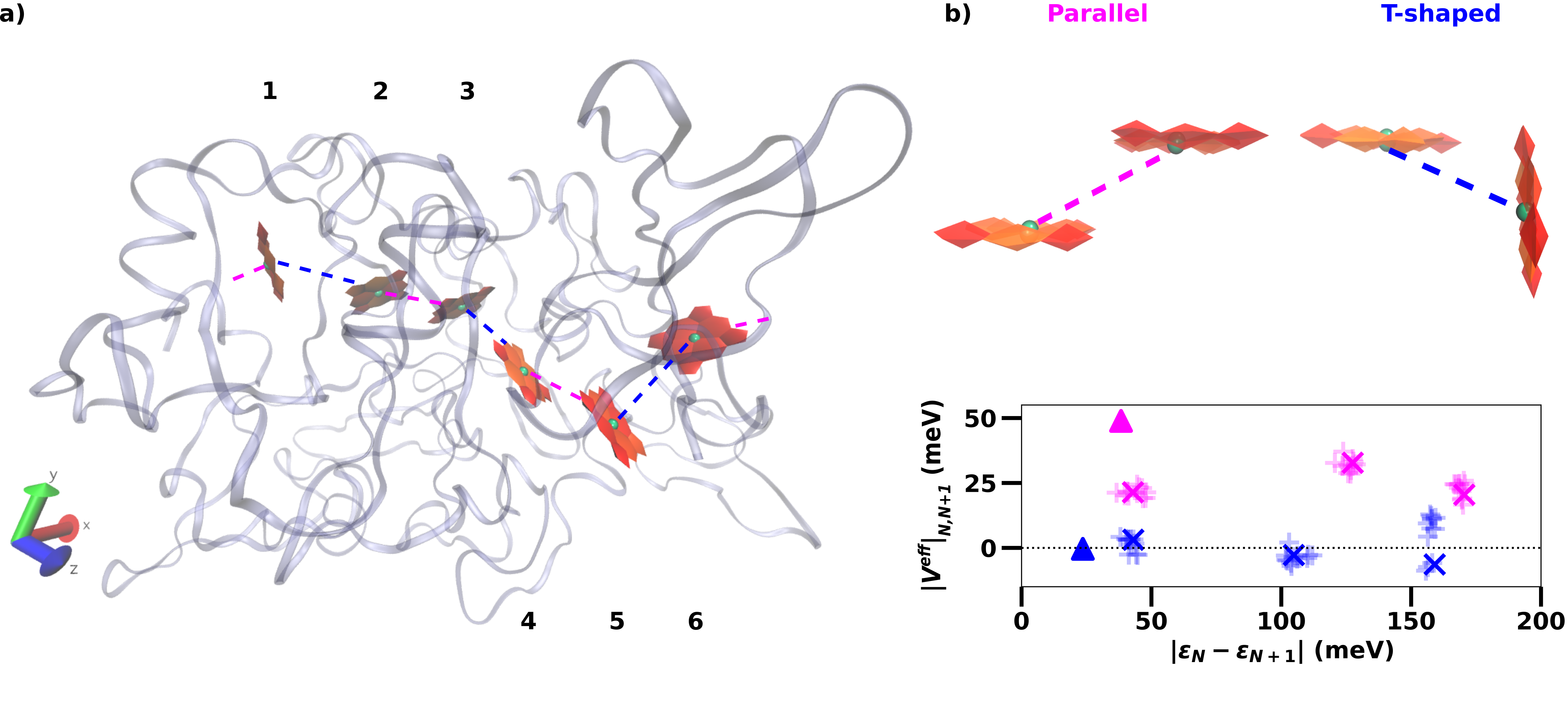}
    \caption{a) The configuration of heme molecules (labeled 1$-6$) studied in this work, shown along the protein backbone of OmcS (plotted as ribbons) extracted from Ref.~\citenum{OmcS}. b) (Top) An example of two heme molecules in the parallel and  T-shaped configuration. (Bottom) Calculated $|V^{eff}|_{N,N+1}$ and $|\varepsilon_N - \varepsilon_{N+1}|$) for six possible heme pairs of OmcS (labeled as X). $2^{\circ}$ rotations of each heme along the Cartesian axes (+) and geometry-optimized structures in implicit solvent (triangles) are also shown.}
    \label{fig:7hemes}
\end{figure*}

OmcS contains six repeating heme molecules arranged along a protein backbone (Figure~\ref{fig:7hemes}a). We compute site energies and coupling for all pairs within unrestricted DFT using the hybrid functional of Perdew, Burke, and Ernzenhof (PBE0)\cite{PBE0} as implemented in the Turbomole Package.~\cite{turbomole2020} With the positional uncertainty of heme orientation in the experimental data,\cite{OmcS} we consider small ($ 2^{\circ}$) inter-molecular rotations around the iron center, which we interpret as accounting for thermal motion. As shown in Figure~\ref{fig:7hemes}b, the heme molecules span $\sim$180 meV in $\varepsilon_N$, reflecting a wide range of static disorder in nuclear geometries, consistent with projector-based methods.~\cite{Dahl_SciAdv2022,Blumberger_JPCL2020,Guberman-Pfeffer_2022,Guberman-Pfeffer_2023} The inter-heme coupling strengths $|V^{eff}|_{N,N+1}$ range from $\sim 0 - 30$ meV, indicating a weak coupling regime consistent with diffusive conductivity.  Our predicted coupling values are in qualitative agreement with prior studies (See SI Section 3).~\cite{Blumberger_JPCL2020,Dahl_SciAdv2022,Guberman-Pfeffer_2022,Guberman-Pfeffer_2023} As expected,\cite{Dahl_SciAdv2022,Blumberger_JPCL2020,SmithJPCB_2006,Guberman-Pfeffer_2022,Guberman-Pfeffer_2023} coupling is highly dependent on inter-molecular orientation: T-shaped dimers show relatively weak coupling ($V^{eff}\sim 0$-$10$ meV), while the parallel pair configurations are more highly coupled ($V^{eff}\sim 20$-$30$ meV). In addition, small inter-heme rotations result in changes of up to  $\sim 20$ meV in site energy misalignments and $\sim 10$ meV in coupling (see Figure~\ref{fig:7hemes} and SI Table S6).

Figure~\ref{fig:7hemes} also includes the coupling and site energy misalignments for the DFT-predicted energy minimized structures of the heme pair in implicit solvent \textcolor{black}{where} excluding the constraint of the protein backbone on the heme \textcolor{black}{leads to drastic changes}. For the energy minimized parallel heme pairs, the geometry is close to the cryo-EM structure, but there is a slight reduction in the distance between hemes resulting in a nearly two-fold increase in the effective coupling (see SI Figure S5). On the other hand, for the T-shaped pairs, the coupling remains weak due to lack of spatial overlap between sites while the energy minimized structure significantly differs from measured pairs\cite{OmcS}.

\subsection{Variation of conductivity parameters with inter- and intra-heme distortions}
Due to their structural softness, thermal vibrations within hemes occur at low frequencies, e.g., 
inter-heme motion has $\omega \sim 10^{12} Hz$ and intra-molecular motion has $\omega \sim 10^{12}-10^{13}$Hz. Using molecular mass $m \sim 500-1000$ amu and considering a simple model of Harmonic phonons, $\frac{1}{2}m \omega^2x^2 = k_BT$, the displacement ($x$) can easily reach $0.5~\textrm{\AA}$ at room temperature. Such large structural distortions may significantly alter $\varepsilon_N$ and $V_{eff}$ in our model, and hence strongly impact conductivity.

\begin{figure*}[htbp]
    \centering
    \includegraphics[width=0.7\textwidth]{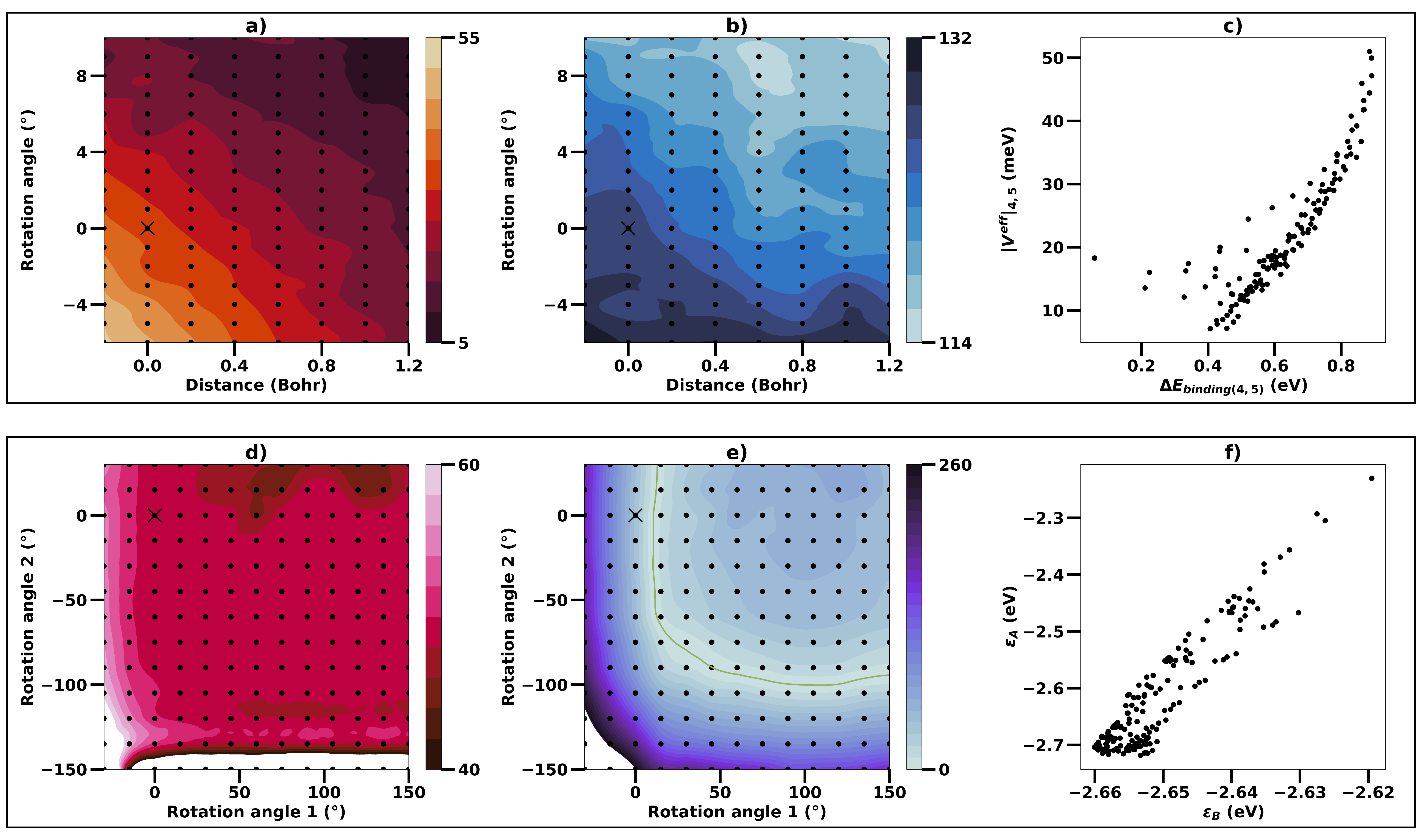}
    \caption{The impact of (a-c) inter-molecular and (d-f) intra-molecular distortions on DFT-predicted conductivity parameters. (a) Effective coupling and (b) site energy differences  upon modifications of inter-heme distance and rotation for the type 4-5 heme pair. The original geometry is marked with an x and data points are shown as gray points. (c) The
    effective coupling as a function of dimer binding energy, showing a positive correlation between the two parameters. (d) Effective coupling and (e) absolute site energy differences upon rotation of the propionate group on one molecule (molecule 1) with respect to the other (molecule 2) for the geometry-optimized heme structure. (h) The site energy for the heme molecule containing the distortion $\varepsilon_A$ compared with the other molecule ($\varepsilon_B$). The isosurface plots were interpolated using a cubic spline method to create a smooth surface as described in the SI.}
    \label{fig:distortions_and_parameters}
\end{figure*}

To examine how the calculated conductivity parameters are affected by a finite inter-heme structural distortion (See SI Fig. S6 for the structure), we systematically modify the orientation of a parallel dimer configuration (types 4 and 5 in Figure~\ref{fig:7hemes}) as an example. Keeping the intra-heme geometry constant, we modify the Fe-Fe distance and change inter-heme orientation by rotating the type-4 heme around the Fe atom in its $yz$ plane. The effective coupling significantly increases as the distance between hemes is decreased or the porphyrin rings of the heme are rotated closer together as shown in Figure~\ref{fig:distortions_and_parameters}a). While coupling can be increased by changing inter-heme structure, this increase may come at a small cost of site energy misalignment (as shown in Figure~\ref{fig:distortions_and_parameters}b). Notably, the relationship between misalignment and coupling is not linear and there are pockets of minimal site energy misalignments and higher coupling (e.g., the point -4 $^{\circ}$, 0.0 \AA). 

Interestingly, we find a correlation between binding energy and the effective coupling for this heme pair as shown in Figure~\ref{fig:distortions_and_parameters}c. This is due to the fact that the charge carrier site overlap and van der Waals interactions that dominate binding are similar functions of inter-heme nuclear geometry (Table S5). This also holds when comparing the different heme pairs of OmcS as shown in Section S5 of the SI. Thus, for parallel heme dimers, energetically favorable inter-heme geometry correlates with larger effective coupling for electron transport. We note that geometry optimization of the peptide pair in implicit solvent leads to a stronger binding energy between hemes as well as larger coupling because the heme pair is no longer constrained by the polymer environment; in other words, the protein environment restricts the coupling of the heme pair. 

Next, we consider the significance of intra-heme \textcolor{black}{distortions} on conductivity parameters. In particular, we focus on the rotation of the charged propionate groups on the heme molecule (see Fig. S7 for the structure). These groups can move relatively freely as rotations around the carbon-carbon single bonds are unhindered and thus, we expect that such rotations will occur freely in the protein environment. An indicator that these propionate rotations could occur is the poor resolutions of some propionate groups in the cryo-EM measurements.~\cite{OmcS} We hypothesize they will result in modification of the electrostatic environment around the heme and therefore can significantly impact site energies. The impact of such rotations is presented in Figure~\ref{fig:distortions_and_parameters} d-f. \footnote{We note that because we freeze all other atoms in the molecule, rotation can cause a steric collision between the propionate groups and so some rotations become energetically unfavorable in our calculations. We focus on the more energetically favorable rotations in Figure~\ref{fig:distortions_and_parameters}.} 
Propionate rotations do significantly change site energy misalignment and have a negligible impact on inter-site coupling. Focusing on rotations that are within 0.1 eV of the minimum energy configuration, the coupling varies over $10$ meV, while site energy differences vary by up to $60$ meV. The site energy change is mainly on the molecule that contains the distortion as shown in Figure~\ref{fig:distortions_and_parameters}f. This is because the electrostatic field created on that molecule is largely screened out on neighboring molecules; i.e., the electrons on the heme molecule alone are sufficient to screen the motion of the charged group from the neighboring heme molecule. For a larger energy cost, the inter-site coupling varies over $\sim 50$ meV while site energy differences range over $\sim 370$ meV, notably inverting the energy ordering with the site energy difference as shown by the green line in Figure~\ref{fig:distortions_and_parameters}d. 

Overall, Figure~\ref{fig:distortions_and_parameters} shows that charge carrier site energy misalignments and site couplings can be tuned to a certain extent by modifying the heme geometry in the parallel configuration: these charge carrier parameters would vary under nuclear geometry distortions due to molecular vibrations. This study also reveals that by modifying the electrostatic environment of one heme with respect to the other \textcolor{black}{through local charged groups}, site energy differences can be modified, suggesting a synthetic route to tuning the roadblock to conductivity in these systems.

 \subsection{DFT-informed conductivity modeling}
\begin{figure*}[htbp]
    \centering
    \includegraphics[width=0.6\textwidth]{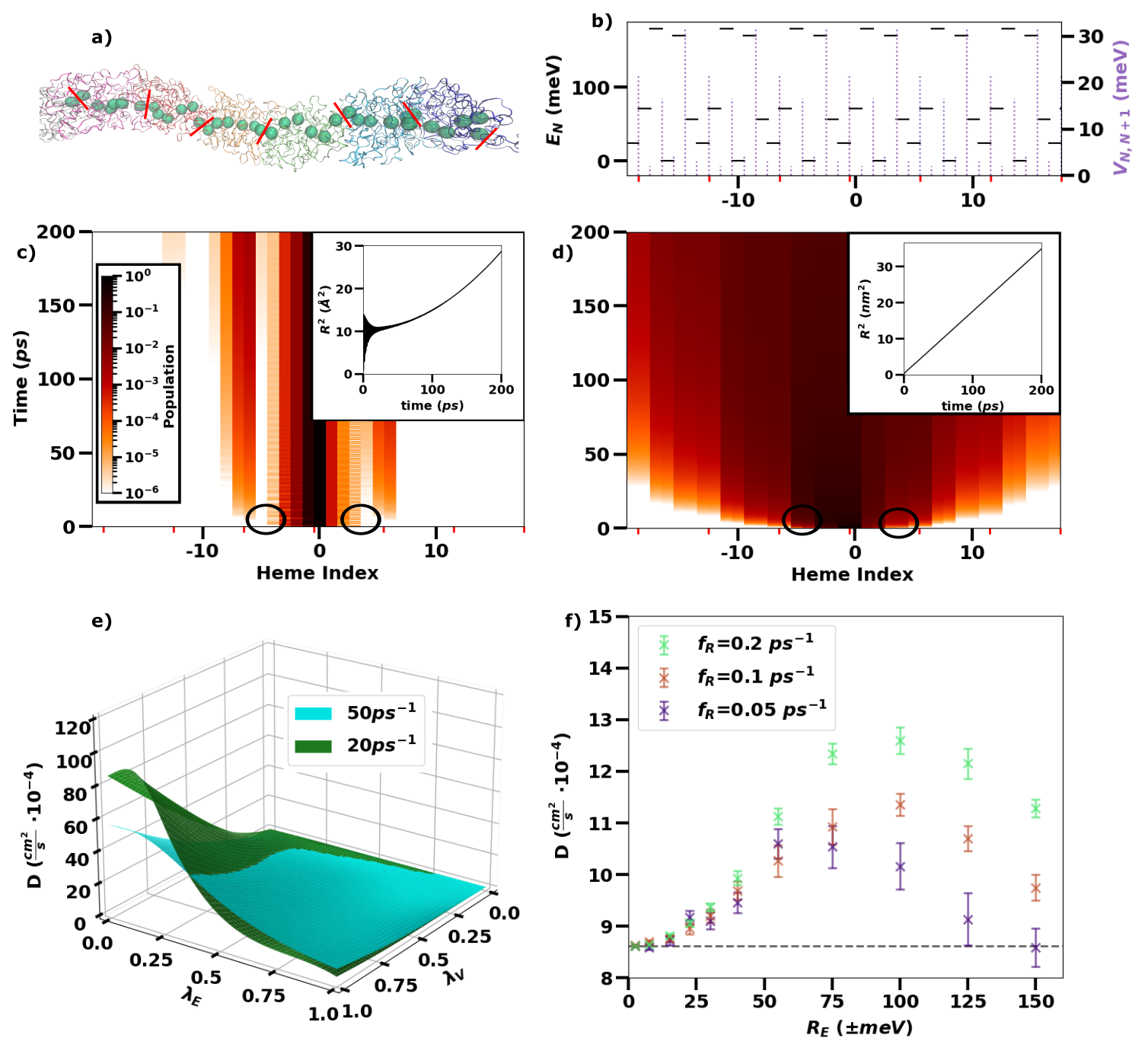}
    \caption{(a) Diagram of repeating heme molecules where the red dash indicates the end of a six-heme unit. (b) Site energy as a function of heme index. Site 0 indicates the initial electron position on heme type 1. (c) Population dynamics without dephasing ($\gamma$ = 0 $ps^{-1}$) with the wavefunction spread $R^2(t)$ as an insert. (d) Population dynamics in the presence of dephasing with $\gamma$ = 50 $ps^{-1}$with the wavefunction spread $R^2(t)$ as an insert. (e) The diffusion coefficient as a function of site energy difference and electronic coupling scaled by $\lambda_{E}$ and $\lambda_{V}$, respectively. Two different dephasing parameters, $\gamma$, are shown.(f) Impact of randomization and rerandomization rate (f$_{R}$) on the diffusion coefficient (D) for $\gamma$ = 50 $ps^{-1}$. f$_{R}$ represents the time before site energies are randomized again.}
    \label{fig:carrier_diffusion}
\end{figure*}

 To model carrier diffusion, we build a chain with 20 segments with each segment containing the six hemes of OmcS with the Hamiltonian built from the DFT-computed site energies and couplings. Following Ref.~\citenum{RZB_model}, we initiate the charge carrier on a given molecular site (Heme Index = `0' in Figure~\ref{fig:carrier_diffusion} b) and compute propagation of the density matrix, $\rho(t)$, across the chain. The conductivity is proportional to the mean square displacement (MSD) along the chain, denoted as $R^2$(t), 

\begin{equation}
   R^2(t) = \mathrm{tr}\bigg\{\left(\mathbf{r}_z \right)^2\boldsymbol{\rho}(t)\bigg\} 
   -\left(\mathrm{tr}\bigg\{\mathbf{r}_z \boldsymbol{\rho}(t)\bigg\} \right)^2,
   \label{eqn:MSD}
\end{equation}

where $\mathbf{r}_z$ is the position operator in the direction of the chain. 

We first explore the coherent density matrix propagation over an energetic landscape as shown in Figure~\ref{fig:carrier_diffusion} b according to the one-dimensional von Neumann equation,
 $\frac{d\rho}{dt} = \frac{-i}{\hbar}[H,\rho]$. As the carrier is initially localized, the density matrix expands over different eigenstates of the system. In the absence of dynamic disorder, we observe Rabi oscillations of the density (Figure~\ref{fig:carrier_diffusion}c), as widely observed for quantum systems.\cite{Mukamel-textbook-hibert-chapter} Because the initial site is in an energy well, it behaves as a trap for the electron carrier, resulting in density oscillations for up to $t \simeq 50$ ps before the wave packet spreads across an extended range (see inset of Figure \ref{fig:carrier_diffusion}c). We find that these initial damped oscillations are present when the carrier is initialized on other low energy sites such as heme types 3 and 6 (see SI section 8). 

Dynamic disorder for quantum states in the heme chain via thermal fluctuations, the environmental coupling, and phonons, should cause occasional dephasing of the charge carrier wave function. To study the impact of dynamic disorder, we introduce quantum mechanical dephasing at a rate $\gamma$,\cite{RZB_model} which reduces the coherences in the density matrix of the Linblad term (Figure \ref{fig:carrier_diffusion}d-~\ref{fig:carrier_diffusion}e). For the dephasing rates considered here, the Rabi oscillations subside at $\sim 1 ps$. As shown in Figure \ref{fig:carrier_diffusion}d, a chosen $\gamma$ = 50 $ps^{-1}$ effectively suppresses the t $\sim 0$ oscillations in the density matrix, and leads to faster spread of $\rho$ than when $\gamma$ = 0 $ps^{-1}$, indicating enhanced diffusion with dephasing. As the charge carrier sites are brought out-of-phase through dynamic disorder, the large site energy differences that limit diffusion can be overcome. Interestingly, the rate-determining step (circles in the plots) with and without dephasing occurs at the same site (between hemes 3 and 4 for leftward diffusion and heme types 5 and 6 for rightward diffusion). Overall, the inclusion of dephasing seemingly enhances the efficiency of electron transport, regardless of the initial carrier location. Thus, \textcolor{black}{in agreement with expectation,} these calculations show the rate-limiting step for diffusion occurs when adjacent hemes exhibit either a large difference in on-site energy or weak inter-heme coupling, with the first factor being more critical than the second because of the large static disorder among hemes in OmcS.

It is important to note that in our case, the wave packet remains within the 120-molecule chain, and D continues to be a linear function of time. Therefore, the choice of boundary condition (e.g., open-end boundary in our study or periodic boundary as used by some authors in the literature~\cite{Song_2015,Yan_2019,Kloss_2019,Cioslowski_2024}) does not significantly affect the results. However, if the simulation time were extended, the wave packet would eventually reach the boundary, leading to reflection and self-interference, rendering the conductivity model invalid for systems with a limited number of molecules.

The diffusion coefficient, calculated as the slope of the MSD after an adequately long time,

\begin{equation}
    D = \frac{1}{2} \frac{R^2(t_{2})-R^2(t_{1})}{t_{2}-t_{1}} \,,\,
    t_2 >> t_0 \,,\, t_1 >> t_0,
    \label{eqn:diffusion}
\end{equation}

\noindent
is 2.5$\times$10$^{-4}$cm$^2$/s for a small dephasing rate of $\gamma$=10 ps$^{-1}$, while at larger $\gamma$ = 100 ps$^{-1}$, it increases to 1.25$\times$10$^{-3}$cm$^2$/s. For the zero dephasing cases, D is of the order of 10$^{-6}$-10$^{-5}$ cm$^2$/s for all initial carrier locations (see SI Section 8).

We next consider the influence of the magnitude of coupling and site energy differences on charge diffusion. We introduce two prefactors, $\lambda_E$ and $\lambda_V$, to scale the DFT site energy differences and couplings $\varepsilon_{N}$ and $V^{eff}_{N,N+1}$, respectively, going from zero to one in the Hamiltonian (Figure~\ref{fig:carrier_diffusion}e). $D$ maximizes for $\lambda_E \sim 0$ (energy resonance between all hemes) and $\lambda_V$ = 1, which represents a scenario mirroring diffusion within a quasi-uniform and a highly coupled heme chain. Near these scaling factor values, an increase in the dephasing rate decreases the diffusion coefficient because dynamic disorder perturbs the homogeneity of the chain's site energies. On the other hand, there exists a contour line on the $D$-$\lambda_E$-$\lambda_V$ surface across which the effect of dephasing transitions from hindering to improving diffusion. This is further illustrated in SI Figure S11, where one parameter (either $\varepsilon_{N}$ or $V^{eff}_{N,N+1}$) is fixed while the other is varied continuously. When $\lambda_V$ is fixed, dephasing becomes more important for facilitating charge carrier diffusion as $\lambda_E$ is increased. Conversely, when $\lambda_E$ is fixed to a non-zero value, dephasing always enhances charge carrier diffusion. Thus, site energy differences represent the bottleneck to carrier diffusion in OmcS, which can be effectively overcome by dephasing and dynamic disorder due to phonons.

While the Lindblad dephasing term includes the \textcolor{black}{frequency-dependent} perturbative impact of phonons, it does not include non-perturbative vibrations, which would modify the electronic Hamiltonian over a large range. As the range of variation in site energy and inter-site coupling in Figure~\ref{fig:distortions_and_parameters} are comparable to expected phonon-induced renormalization, on the order of $\sim 100$ meV, the effect of molecular vibrations on conductivity should not be treated simply as a perturbation. Instead, it necessitates consideration at the level of strong and frequent alteration of the Hamiltonian. To properly account for this factor, we go beyond Ref.~\citenum{RZB_model} to incorporate non-perturbative intra-heme vibrations by introducing a randomization term in the Hamiltonian $H$ such that the site energies can be adjusted with a frequency $f_R$ at a given time $t$ bound by a value $R_{E}$ (i.e., $  \varepsilon_N \rightarrow \varepsilon_N + R_{E}\cdot\xi_{N}(t|f_R)$ in Eq.~\ref{eqn:H_lambda}). 
The term $\xi_{N}$ is a random number that scales the re-randomization energy every time-step and is recalculated according to the randomization rate. Indeed, our DFT calculations show the sensitivity of site energies to the motion of dipolar groups on the heme molecule as shown in Figure~\ref{fig:7hemes}, suggesting that readjusting  $\varepsilon_N$ is necessary and may be important for the study of conductivity through OmcS in physiologically relevant conditions. Physically, $f_R$ is the frequency associated with inter-molecular phonons which are presumably in the THz or sub-THz regime. Thus, we  introduce randomization in $\varepsilon_{N}$ at different rates, i.e., every 0.5 ps $-$ a time scale representative of high-frequency intra-molecular vibration of light ligands or equivalently optical phonons $-$ or every 20 ps $-$ representative of fluctuations from low-energy acoustic phonon modes).

Figure \ref{fig:carrier_diffusion}(f) depicts the impact of such randomization on the evolution of the density matrix, with conductivity parameters from DFT. As expected, minor randomizations with small $R_E$ do not affect the diffusion rate because the overall `roughness' of the energy landscape remains unaltered, and the wave packets are still confined by the deep energy wells depicted in Figure \ref{fig:carrier_diffusion}(b). When $R_E$ is increased, however, nonlinear increases in $D$ are observed, especially with higher rates of randomization. $D$ saturates and gradually decreases as $R_E$ passes 70 meV since the randomization enlarges site energy differences and hinders diffusion. We note that while these non-perturbative dynamic changes to the Hamiltonian can significantly alter the conductivity, frequency-dependent perturbative dynamic disorder introduced through $L$ have a much more significant impact. 

Considering a semi-classical model of electron transport, the diffusion constant, $D$ is related to mobility $\mu$ by the Einstein relationship (D = $\mu$k$_B$T/q). Using this formulation, we predict $\mu \sim 10^{-4}$-10$^{-5}$ cm$^{2}$/Vs without the inclusion of dephasing (fully coherent transport), similar in magnitude to that reported for T = 270 K in Ref.~\citenum{Dahl_SciAdv2022}. With the inclusion of dephasing, within the dephasing rates considered in this work \textit{D} can increase by 4-6 orders of magnitude, reconciling with experimentally measured values.

 \section{Discussion}

Our study highlights a new perspective on the importance of nuclear motion and phonons in heme protein conductivity. It is known that thermal vibrations impact electron transport through multi-heme \textit{c}-type cytochromes via their impact on site energies/reaction barriers and couplings. By applying a quantum transport model that includes both electronic coherence and dephasing, we demonstrate the impact of dynamic, incoherent, frequency dependent modulation of sites via thermal fluctuations. We hypothesize the underestimation of predicted conductivity for OmcS by more than 10,000-fold~\cite{Guberman-Pfeffer_2022, Dahl_SciAdv2022,Blumberger_JPCL2020} may be due to this dynamic disorder, and we show that incorporating proxy effects of phonons in a conductivity model indeed increases calculated charge carrier diffusion constants. Specifically, frequency dependent Lindblad dephasing, describing perturbative electron-phonon interactions, enables electrons to overcome wavefunction localization in the charge carrier diffusion model, and site energy fluctuations, representing strong electron-phonon interactions, also increases carrier diffusion constants (Figure~\ref{fig:carrier_diffusion}). We note this model is complementary to Marcus theory, where the electron strongly (and non-perturbatively) couples to nuclei, resulting in \textcolor{black}{reorganization} of the molecule, in order to transfer between sites.

Our results suggest distinct regimes of site energy misalignment, from perfect to random alignment, within which this dynamic disorder can have opposing effects. This phonon-assisted charge diffusion is a mechanism suited to characteristically biological, disordered systems; in a perfectly ordered material with aligned heme site energies, dynamic disorder diminishes charge carrier diffusion (Figure~\ref{fig:carrier_diffusion}e). Because the environment around each heme molecule in the naturally occurring OmcS polymer varies greatly, the molecular structure of each heme and hence site energies vary significantly as well. Thus, we predict OmcS lies within the regime of phonon-assisted charge diffusion. 

Our results can also be interpreted within the framework of the transient (de) localization model,\cite{Fratini_2016,Flickering_Resonance,Giannini_2019,Ciuchi_2011,Sneyd_2022} whereby the presence of phonons can dynamically impact the delocalization of the charge carrier state and therefore impact carrier conduction. In our study, both dephasing and randomization effects are shown to lead to a noticeable increase in conductivity when the energy landscape is rough across the molecular chain (e.g., $\lambda_E >$ 0.4 in Fig. 4e and the monotonic increase of $D$ for R$_E <$ 75 meV in Fig. 4d). Different rates of decoherence and randomization processes are directly linked to how quickly charges localize or delocalize. The critical role of dephasing in overcoming the \textcolor{black}{scattering} of wave packets can be understood as thermally induced delocalization. In that sense our model agrees with transient (de)localization in that site energy fluctuations can contribute to diffusive transport behavior, in both near- and non-resonant site energy cases. With further development of DFT approaches for predicting dynamic parameters, we believe that the use of the Lindblad master equation may quantitatively model electronic transport through complex materials, incorporating environmental and thermal effects.

The magnitude of site energy randomization (Figure~\ref{fig:carrier_diffusion}f) is supported by the range of energies accessible by effects of intra-heme molecular \textcolor{black}{motions}. Rotation of propionate groups, for example, can significantly affect the site energies (Figure~\ref{fig:distortions_and_parameters}e) and are of sufficient magnitude to flip the sign of the difference in site energies between heme pairs. The total range over which each site energy randomizes is about +/-100 mV at the maximum increase in diffusion constant (Figure~\ref{fig:carrier_diffusion}f), comparable with computed temperature dependent changes in site energies.~\cite{Dahl_SciAdv2022,Guberman-Pfeffer_2022} While not resolved in structural studies of OmcS,~\cite{OmcS} control over the conformations of heme porphyrin substituents may be a novel biological mechanism for dynamic modulation of the charge transport free energy surface of cytochrome wires.

Lastly, we note that in comparison to energy minimized structures of heme pairs in implicit solvent, heme pairs in the OmcS cryo-EM structure are not optimized for coupling energy, and they are offset from free energy minima (Figure \ref{fig:7hemes}b). This result suggests that this highly conserved,\cite{Baquero_2023} one-dimensional heme packing motif is evolved under constraints coincident with evolutionary pressure to facilitate extracellular electron transfer. The protein is holding the heme pairs in positions that are not functionally optimized for electron transfer nor represent heme configurations at free energy minima. In light of the role of significant electron-phonon interactions that overwhelm static site energy misalignment and low coupling energies (Figure~\ref{fig:carrier_diffusion}d-f), one possible interpretation is that this arrangement is the most dense packing of heme achievable within biosynthetic limitations, such as those imposed by heme maturation processes or simply the genetically regulated control of secreted heme wires. Given the beneficial impact of both perturbatively and non-perturbatively varying the charge carrier parameters discussed in the previous section, the sub-optimal binding and coupling of the heme dimer may be the price of varying charge carrier parameters for improved transport in these biological nanowires.

\section{Conclusions}
In this article, we introduced an approach that computes trends in the effective coupling and site energies associated with charge carrier sites and applied this approach to the natural conducting protein, OmcS. With a first-principles-based parameterization of the model of Ref.~\citenum{RZB_model}, the predicted parameters provided input to an electron carrier Hamiltonian, which together with an empirical constant dephasing rate resulted in a range of predicted charge carrier diffusivity that can reach the order of magnitude of experimental conductivity. By introducing randomization, a parameter to describe non-perturbative phonons, we inferred that intra-molecular vibrations can also allow the system to overcome static disorder. Additionally, we demonstrated the ability to modify the calculated conductivity parameters by inter- and intra-molecular distortions. Through this analysis, we determined that the intra-heme electrostatic environment significantly impacts, suggesting a route to synthetic control of static disorder, the bottleneck to conductivity in OmcS. Our study indicates that site energy disorder among the heme molecules results in non-resonant, phonon-enhanced transport, which differs fundamentally from the typical behavior in organic crystals with minimal static disorder.

\section{Materials and Methods}
\subsection{OmcS structure}
The structure of OmcS was obtained from nanowire filament PDB structure 6ef8, measured via cryo-EM (see Figure~\ref{fig:7hemes}).\cite{OmcS} Hydrogen atoms were added using VMD 1.9.3 and the psfgen 2.0 plugin\cite{HUMP96} for the CHARMM36m force-field and the 2013 parameterization of the heme type C group.\cite{HEC_FF,CHARMM36m_FF}
We report the protein residue labels of the group in the PDB files, with heme type C ``residue" labelled as HEC. The heme pair subsystems were truncated at the $\beta$-carbon of the deprotonated-Cysteine (CYC and CYD) and unprotonated-Histidine (HSD) residues of the heme group, replacing the $\alpha$-carbon with Hydrogen following previous studies.\cite{Sulphur_needed_2017,Spin_Dependent_coupling}
The propionate groups of HEC were left deprotonated to mimic the biological 
pH-buffered environment, without the inclusion of explicit water solvent molecules given that 
cryo-EM showed no structured water near these group. Iron is set to the low-spin Fe(III) state, which we expect will occur in this coordination environment.\cite{Walker_low_spin_hemes,Spin_Dependent_coupling} 

Individual heme molecules were extracted to determine the site energies, and heme pairs extracted for calculating coupling. In this extraction to give parameters for all charge carrier sites, all heme  groups from Chain A of 6ef8 are extracted, along with the first heme group of Chain C. 
Noting the limited resolution of the cryo-EM study,\cite{OmcS} we create a range of snapshots by rotation of heme molecules by  2 degrees around the iron center, along the Cartesian planes, thus sampling our parameters in order to account for experimental resolution (and thermal fluctuations). 

\subsection{DFT computational details}
All-electron unrestricted-spin density functional theory (DFT) calculations were performed on the heme molecules and pairs with the nuclear geometry of all atoms, except hydrogen, directly from cryoEM fits, while hydrogen positions were optimized using classical force fields as described above. We expect that these structures provide a representation of an average room temperature conformation of the heme and protein.

DFT calculations were performed in \textsc{Turbomole}7.6,\cite{turbomole2020} using the hybrid functional, PBE0,\cite{PBE,PBE0,otherPBE0} with a grid size of four for determining the gradient of the density, and the def2-SVP energy-consistent double $\zeta$ basis set for all atoms.\cite{def2-basis} We note that PBE0 functional has been found to be sufficient for describing the electronic structure of individual porphyrins\cite{Por21_DFT_benchmark}, which are chemically similar to our heme molecules, and iron-containing phthalocyanines\cite{Marom_2009}. To describe the natural conditions of the heme protein and stabilize the overall negative charge of the heme groups, the implicit solvent model COSMO was applied using standard parameters for water; with dielectic constant 78.390.\cite{cosmo-paper} We note that the role of explicit solvent may be important particularly for describing the electronic structure in the presence of charged residues~\cite{ACC-Hex} and we may expect the lack of this description will quantitatively impact the molecular energetics. Grimme's D3 dispersion correction was added to the total energy for binding energy calculations.\cite{disp3} The resolution of identity approximation for the Coulomb integral (RI-J) was used to speed up the calculation.\cite{RI-J_paper} The total energy convergence criteria was set at 10$^{-7}$ Hartrees, and the one particle reduced density matrix convergence criteria was set at 10$^{-7}$. In the Fe(III) oxidation state there is one unpaired electron localized on each heme group. The unpaired spins between hemes was set to parallel and anti-parallel in separate calculations by fixing the number of alpha spin and beta spin electrons. For geometry optimization, forces were converged to 5 $\times$10$^{-3}$ meV/\AA{}.

We found that for most Fe(III) heme molecules, the lowest energy unoccupied molecular orbital (LUMO) is the charge carrier site, aligning with expectations, with the exception of heme type 1. For dimers composed of these systems (6+1 and 1+2), the LUMO+1 state corresponds to the isolated monomer LUMO character (based on the presence of Fe $3d_{yz}$ character in the $\pi$-orbital). Thus, we consider the LUMO+1 of heme 1 (with smaller iron $3d_{yz}$ character) as the charge carrier site. Additionally, because of the small coupling in the T-shape, some calculated couplings were imaginary numbers. In order to ensure all coupling values used in the conductivity model are real numbers, hemes 2 and 3 were slightly modified with respect to the pdb file\textcolor{black}{, within the resolution of the measured electron density}. Heme 2 was rotated by -2 degrees in the xz plane and heme 3 is rotated +2 degrees in the xy plane, while all other hemes retain the 6ef8 geometry.

The heme pair binding energy was computed as 
\begin{align}
   \Delta E_{binding} = E_{1,2} - E_1 - E_2, 
\end{align}
where $E_{1,2}$ is the total energy of the interacting pair and the E$_{1/2}$ are the total energy of the isolated heme molecules 1 and 2.

\subsection{Conductivity modeling details}
For conductivity modelling, we set up the heme wire basis as a polymer of 20 6-heme chain units, where each heme chain starts with heme type 1 (506 in the pdb) and ends with heme type 6 (505 in the pdb). The number of repeating units is chosen to be long enough such that the boundary or ends of the chain do not influence charge carrier propagation. Note that the ends of the wire are not joined and the wire is not periodic. We index these heme charge carrier sites from $N = -60$ and increase the index in the up-chain direction until $N = 60$. SI Table S4 reports the corresponding position operator in the z-direction for this system for $N = 0$ to $N = 6$, based on the 6ef8 chain A structure. In addition to the parameterization described above, there are two heme C monomers on adjacent chains; these two site energies are averaged to obtain the site energy for heme type 506.
The 6-term Runge-Kutta method(RK6) was used to propagate the density matrix in Python3 according to the LvN equation with the Lindblad correction for~\citenum{RZB_model}. The density matrix was initialized with a diagonal value of one on a single site; all other density matrix elements were set to zero at the initial time. The density was propagated with a time step of 1 $fs$, to a total time of 200$ps$ matching Ref.~\citenum{RZB_model}. The randomization frequency ($f_{R}$) can be converted to find the time between new values being added or subtracted to each individual site energy. The charge carrier density matrix is renormalized after every full RK6 propagation step. However, for the range of dephasing rate constants used for this system, it is noted that all oscillations in the MSD subsides around 1$ps$ i.e. the long time limit is reached in around 1$ps$.

\section{Acknowledgments}
This work was primarily supported by the National Science Foundation (NSF) Materials Research Science and Engineering Center program through the UC Irvine Center for Complex and Active Materials (DMR-2011967). SS and LNM acknowledge funding from the NSF CAREER program under grant number DMR-1847774. AIH also acknowledges partial support from DOE Basic Energy Sciences grant number DE-SC0020322. Calculations were run on HPC$^3$ at University of California, Irvine (UCI), which is maintained by UCI's RCIC. Visualization in this paper is carried out in VMD with Tachyon.\cite{HUMP96,STON1998}

\section{Supporting Information}
\textbf{Supporting Information Available:} Theoretical Details; comparison of predicted coupling for \textcolor{black}{decoupling via inter-heme spin flip} with projector methods; comparison of \textcolor{black}{decoupling via inter-heme spin flip} predictions with previous studies of OmcS; information on Rabi oscillations; details of calculated heme pair binding energies; connection of hemes to the 6ef8 PDB file; schematic of inter- and intra-heme distortions considered for Figure~\ref{fig:distortions_and_parameters}; Impact of conductivity model starting site, two degree rotations, and dephasing on diffusion; eigenvalues and eigenvectors associated with the charge carrier.
% Bibliography
\bibliography{ref.bib}

\end{document}